\documentclass[showpacs,fleqn,nobibnotes,twocolumn]{revtex4}

\usepackage{amsmath}
\usepackage{graphicx}

\def\lsim{\raise0.3ex\hbox{$<$\kern-0.75em\raise-1.1ex\hbox{$\sim$}}}
\def\gsim{\raise0.3ex\hbox{$>$\kern-0.75em\raise-1.1ex\hbox{$\sim$}}}

\newcommand{\rk}{\mbox{\boldmath $k$}}

\newcommand{\rr}{\mbox{\boldmath $r$}}

\newcommand{\rb}{\mbox{\boldmath $b$}}
\newcommand{\rx}{\mbox{\boldmath $x$}}
\newcommand{\ry}{\mbox{\boldmath $y$}}

\newcommand{\aaa}{\mathcal{A}}

\newcommand{\bra}{\langle}
\newcommand{\ket}{\rangle}
\newcommand{\N}{\mathcal{N}}

\newcommand{\OO}{\mathcal{O}}

\newcommand{\vi}{,\hspace{-0.065cm}}

\begin{document}

\title{Deeply virtual Compton 
scattering at small $x$ in  future Electron - Ion Colliders}
\author{V. P. Gon\c{c}alves and D. S. Pires}

\affiliation{
High and Medium Energy Group \\
 Instituto de F\'{\i}sica e Matem\'atica, Universidade Federal de Pelotas\\
Caixa Postal 354,  96010-900, Pelotas, RS, Brazil.
}

\begin{abstract}
The study of  exclusive processes in the future electron-ion ($eA$) colliders will be an important tool to 
investigate the QCD dynamics at high energies as they are in general driven by the gluon 
content of the target which is strongly subject to parton saturation effects.  In this 
paper we compute the coherent and incoherent cross sections for the  deeply virtual Compton scattering (DVCS) process relying on the color dipole approach and considering different models for the dipole - proton scattering amplitude. The dependencies of the cross sections with the energy, photon virtuality, nuclear mass number and squared momentum transfer are analysed in detail. We demonstrate that the ratio between the incoherent and coherent cross sections decreases at smaller values of  $Q^2$ and increases at smaller values of $A$. We show that the coherent cross section dominates at small $t$ and  exhibits the typical diffractive pattern, with  the number of dips in the range $|t| \le 0.3$ GeV$^2$  increasing with the mass atomic number. Our results indicate that the position of the dips are independent of the model used to treat the dipole - proton interaction as well as of the center-of-mass energy. 
\end{abstract}

\maketitle

\section{Introduction}

\begin{figure}[t]
\begin{center}
\scalebox{0.35}{\includegraphics{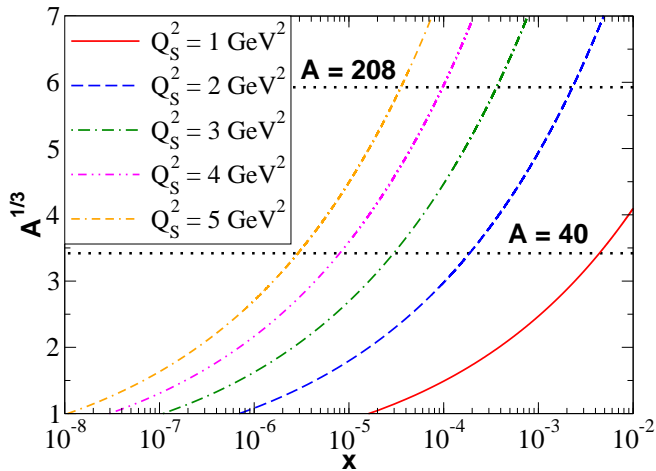}}
\caption{(Color online) Nuclear saturation scale as a function of the Bjorken $x$ and nuclear mass number $A$.}
\label{qsa}
\end{center}
\end{figure}

The study of the hadronic structure in the non-linear regime of the Quantum Chromodynamics (QCD) is one of the main goals of the future Electron - Ion Colliders (EIC) \cite{Raju,Boer,Accardi,LHeC}.  It is expected that the resulting experimental data will be able to determine the presence of gluon saturation effects, the   magnitude of the associated non-linear corrections and 
what is the correct theoretical framework for their description. This expectation is easily understood, if we  analyse the Bjorken-$x$ and nuclear mass number $A$ behaviours of the  nuclear saturation scale, $Q_{s,A}$, which determines the onset of non-linear effects in the QCD dynamics \cite{hdqcd}.     Assuming that the saturation scale of the nucleus is enhanced with respect to the nucleon one by the {\it oomph} factor $A^{\frac{1}{3}}$, it can be expressed by the  parametrization 
$Q_{s,A}^2 = A^{\frac{1}{3}} \times Q_0^2 \, (\frac{x_0}{x})^{\lambda}$. As a consequence  nuclei are an efficient amplifier of the non-linear effects. In Fig. \ref{qsa} we present the theoretical expectations for the saturation scale as a function of $x$ and $A$, considering the  parameters  $Q_0^2 = 1.0$ GeV$^2$, $x_0 = 1.632 \times 10^{-5}$ and $\lambda = 0.2197$ as in Ref. \cite{iims}. We can observe that, while in the proton case we need very 
small values of $x$ to obtain large values of $Q_s^2$, in the nuclear case a similar value 
can be obtained for values of $x$ approximately two orders of magnitude greater.
Consequently, the parton density that is accessed 
in electron - nucleus ($eA$) collisions would be equivalent to that obtained in an electron - proton collider at energies 
that are at least one order of magnitude higher than at HERA.

The enhancement of the non-linear effects with the nuclear mass number  has motivated the development of an intense phenomenology about the implications of these effects in inclusive and diffractive observables which could be measured in the Electron - Ion Colliders \cite{victor,simone1,simone2,Nik_schafer,Kowalski_prl,Kowalski_prc,erike_ea1,erike_ea2,vmprc,simone_hq,Caldwell,vic_erike,Lappi_inc,Toll,Lappi_bal}. These studies indicate 
that the analysis of  inclusive observables, as e.g. the nuclear structure functions $F_2^A$ and $F_L^A$, probably will not be the best way to obtain a signature of the non-linear effects due to the large uncertainty present in the collinear predictions at small $x$. In contrast,
diffractive events in $eA$ collisions are expected to be a smoking gun for the gluon saturation physics. Basically, 
saturation physics predicts 
that in the asymptotic 
limit of very high energies,  diffractive events will be half of the total cross 
section, the other half being formed by all inelastic processes \cite{Nikolaev,simone2}.  
In the kinematical range probed by the future EIC, diffractive processes are expected to contribute by a large amount ($\approx 20 \%$) to the total cross section \cite{erike_ea2}. This observation have motivated  more detailed studies of the diffractive interactions, with special attention to exclusive diffractive vector meson production and the deeply virtual Compton scattering (DVCS), which are experimentally clean and can be unambiguously identified by the presence of a rapidity gap in final state. As these exclusive processes are driven by the gluon 
content of the target, with the  cross sections being  proportional 
to the square of the scattering amplitude,  they  are strongly sensitive to the 
underlying QCD dynamics. In particular, in Refs. \cite{vmprc,vic_erike} we have presented a systematic analysis of these processes in terms of the non-linear QCD dynamics and derived  the energy dependence of the total vector meson ($V = \rho, \, \phi$ and $J/\Psi$) and DVCS cross sections. 
A shortcoming of these studies is that  the distributions on the square of the momentum transfer ($t$), which are an important source of information about the spatial distribution of the gluons in a nucleus and about fluctuations  
of the nuclear color fields, were not estimated. Recently,  these distributions were calculated in Refs. \cite{Caldwell,Lappi_inc,Toll}, where  a comprehensive study about the vector meson production in $eA$ colliders considering different assumptions were presented. Our goal in this paper is to complement these studies by presenting a detailed analysis of the nuclear DVCS (For previous studies see, e.g. Refs. \cite{Kirchner,Freund,Goeke,mag,kope_dvcs,Fazio}). In particular, we update the previous calculations considering the more recent models for the dipole - proton scattering amplitude and presenting  the predictions for the $t$-dependence of the nuclear DVCS cross section.

This paper is organized as follows. 
In the next Section we present a brief review of the description of the nuclear DVCS in the color dipole formalism. In Section \ref{qcd} we discuss the models for the dipole - proton scattering amplitude used  as input in our calculations and present a comparison between its predictions. 
In Section \ref{res} we present our results for the energy, virtuality and momentum transfer dependencies of the DVCS cross sections.  Finally, in Section \ref{conc} we summarize our main conclusions.

\section{Nuclear DVCS in the color dipole picture}
\label{nucDVCS}

\begin{figure}[t]
\begin{center}
\scalebox{0.35}{\includegraphics{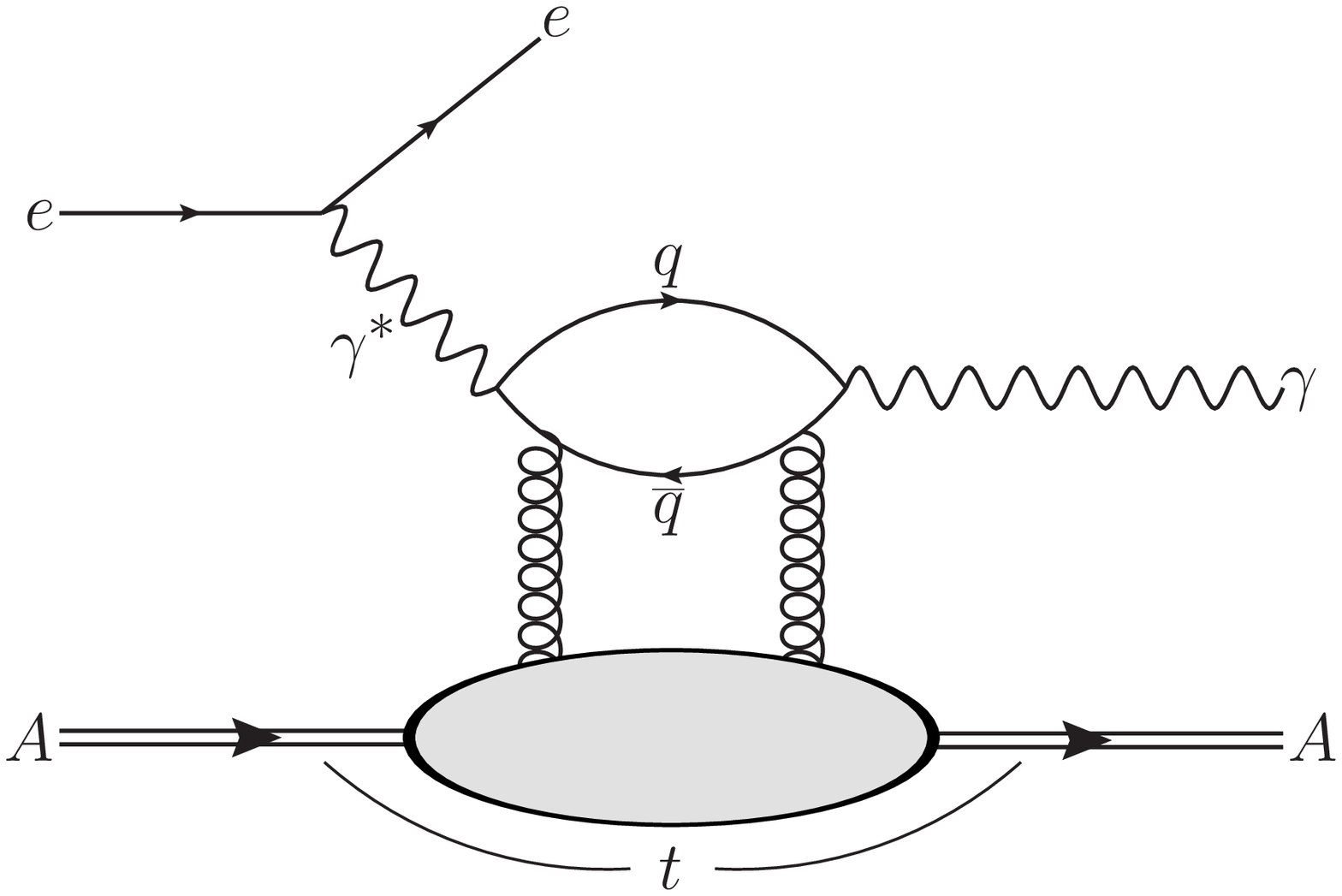}}
\scalebox{0.35}{\includegraphics{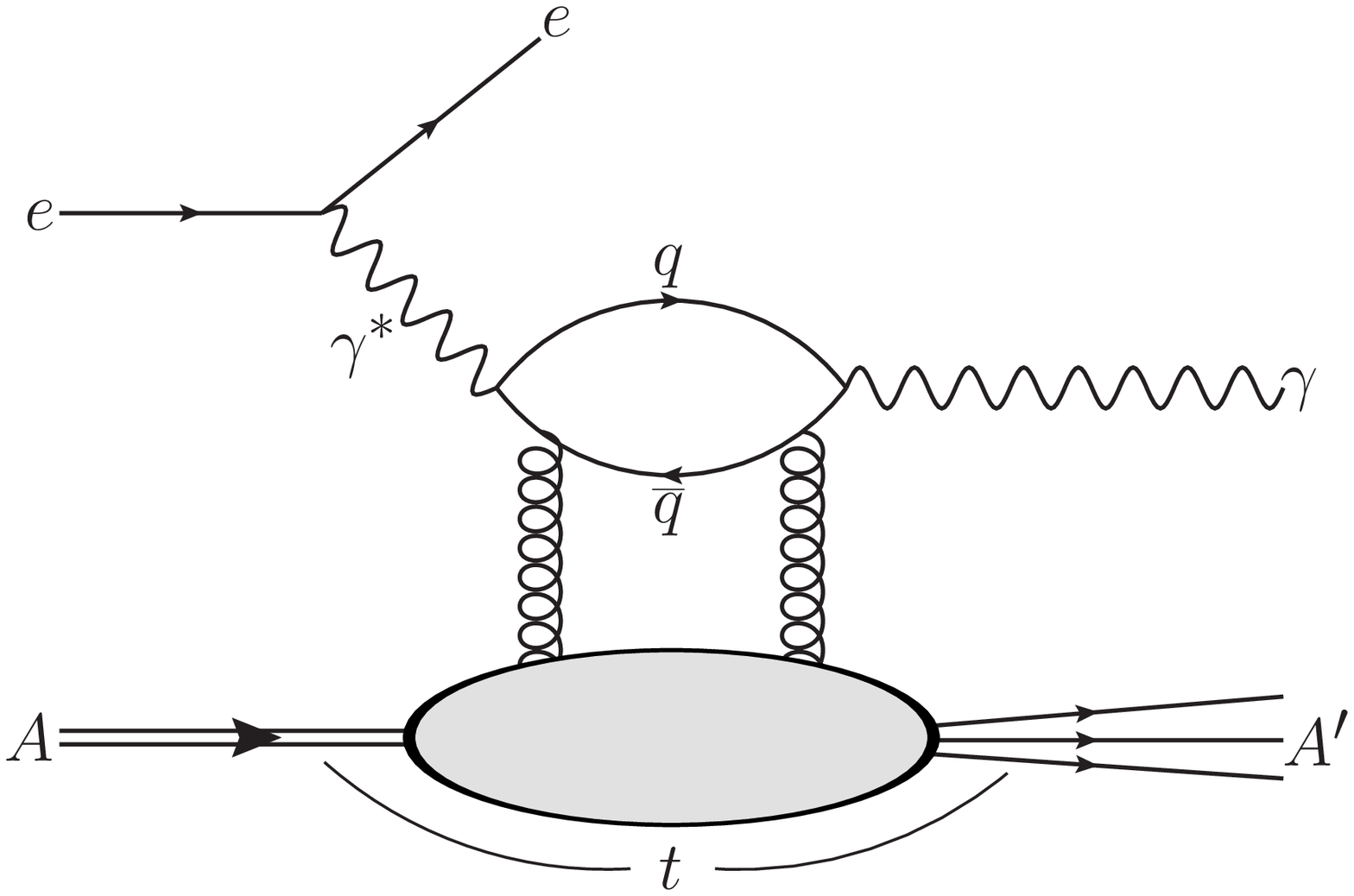}}
\caption{Deeply Virtual Compton Scattering in  coherent (upper panel) and  incoherent (lower panel) $eA$ collisions.}
\label{diagramas}
\end{center}
\end{figure}

Let us start presenting a brief review of the deeply virtual Compton scattering (DVCS) in electron - ion collisions. In the color dipole approach the exclusive  production $\gamma^* A \rightarrow \gamma Y$  in electron-nucleus interactions at high energies can be factorized in terms of the fluctuation of the virtual photon into a $q \bar{q}$ color 
dipole, the dipole-nucleus scattering by a color singlet exchange  and the recombination into 
the exclusive final state $\gamma$. This process is characterized by a rapidity gap in the final 
state. If the nucleus scatters elastically, $Y = A$, the process is called coherent 
production  and can be represented by the diagram in Fig. \ref{diagramas} (upper panel). The corresponding integrated cross 
section  is given in the high energy regime (large coherence length: $l_c \gg R_A$) by 
\cite{vmprc,kop1}
\begin{eqnarray}
\sigma^{coh}\, (\gamma^* A \rightarrow \gamma A)  =  \int d^2\rb \left\langle 
\mathcal{N}^A(x,\rr,\rb) \right\rangle^2
\label{totalcscoe}
\end{eqnarray}
where 
\begin{eqnarray}
\left\langle \mathcal{N}^A \right\rangle = \int d^2\rr
 \int dz  \Psi_{\gamma}^*(\rr,z) \, \mathcal{N}^A(x,\rr,\rb)\, \Psi_{\gamma^*}(\rr,z,Q^2)
 \label{totalcscoe1}
\end{eqnarray}
and $ {\cal N}^A (x, \rr, \rb)$ is the forward dipole-nucleus scattering amplitude for a 
dipole with 
size $\rr$ and impact parameter $\rb$ which encodes all the
information about the hadronic scattering, and thus about the
non-linear and quantum effects in the hadron wave function.
As in our previous studies \cite{vmprc,vic_erike}, 
in what follows we will use in our calculations  the model proposed in Ref. 
\cite{armesto}, which describes  the current  experimental data on the nuclear 
structure function as well as includes the  impact parameter dependence in the dipole 
nucleus cross section (For details see Ref. \cite{erike_ea2}). In this model the forward dipole-nucleus amplitude is given by
\begin{eqnarray}
{\cal{N}}^A(x,\rr,\rb) = 1 - \exp \left[-\frac{1}{2}  \, \sigma_{dp}(x,\rr^2) 
\,A\,T_A(\rb)\right] \,\,,
\label{enenuc}
\end{eqnarray}
where $\sigma_{dp}$ is the dipole-proton cross section and $T_A(\rb)$ is  the nuclear profile 
function, which is obtained from a 3-parameter Fermi distribution for the nuclear
density normalized to $1$. This  equation, 
 based on the Glauber-Gribov formalism \cite{gribov},  
sums up all the 
multiple elastic rescattering diagrams of the $q \overline{q}$ pair
and is justified for large coherence length, where the transverse separation $\rr$ of partons 
in the multiparton Fock state of the photon becomes a conserved quantity, {\it i.e.} 
the size 
of the pair $\rr$ becomes eigenvalue
of the scattering matrix. The differential distribution with respect to the squared momentum transfer $t$ for coherent interactions is given by
\begin{eqnarray}
 \frac{d\sigma_{coh}}{dt} = \frac{1}{16\pi}\left|\aaa^{\gamma^*A\rightarrow\gamma A}(x,Q^2,t) \right|^2,
\end{eqnarray}
where 
\begin{eqnarray}
 \aaa^{\gamma^*A\rightarrow\gamma A}(x,Q^2,t) = 2\int d^2 \rb e^{-i\rb\cdot\Delta}\bra\N^A(x,\rr,\rb)\ket,
\end{eqnarray}
with $\Delta^2 =-t$ and  $\N^A$ given by Eq. (\ref{enenuc}).
 
On the other hand, if the nucleus scatters inelastically, i.e. breaks up ($Y = A^{\prime}$),   
the process is denoted incoherent production and can be represented as in Fig. \ref{diagramas} (lower panel). 
 In this case 
one sums over all final states of the target nucleus, 
except those that contain particle production.  Therefore we have: 
\begin{eqnarray}
\sigma^{inc}\, (\gamma^* A \rightarrow \gamma A^{\prime}) = \frac{|{\cal I}m \, 
{\cal A}(s,\,t=0)|^2}{16\pi\,B} \;
\label{totalcsinc}
\end{eqnarray}
where at high energies ($l_c \gg R_A$) \cite{kop1}:
\begin{eqnarray}
|{\cal I}m \, {\cal A}|^2  =  \int d^2\rb \, A\,T_A(\rb) \left\langle  \sigma_{dp} \, 
\exp\left[- \frac{1}{2} \, \sigma_{dp} \, A\,T_A(\rb)\right]  \right\rangle^2 
\label{totalcsinc1}
\end{eqnarray}
with the  $t$ slope $B$ being  the same as in the case of a 
nucleon target. 
In the incoherent case, the $q\bar{q}$ pair attenuates with a constant absorption cross 
section, as in the Glauber model, except that the whole exponential is averaged 
rather than just the cross section in the exponent.  For the calculation of the differential cross section $d\sigma/dt$ for incoherent interactions we apply for the DVCS process the treatment presented in Ref. \cite{Lappi_inc}. Consequently, we have that 
\begin{eqnarray}
 \frac{d\sigma_{inc}}{dt} = \frac{1}{16\pi}\big\bra\bra |\aaa^{\gamma^*A\rightarrow\gamma A'}(x,Q^2,t)|^2 \ket\big\ket,
\end{eqnarray}
with the scattering amplitude $ \aaa^{\gamma^*A\rightarrow\gamma A'}$ being approximated by \cite{Lappi_inc}
\begin{widetext}
\begin{eqnarray}
 |\aaa^{\gamma^*A\rightarrow\gamma A'}(x,Q^2,t)|^2 = 16\pi^2 B_p^2 &\displaystyle\int& d^2\rb e^{-B_p\Delta^2}\N^p(x,\rr)\N^p(x,\rr^{\prime})\,A\,T_A(\rb) \nonumber\\
  &\times&\exp\Big\{ -2\pi(A-1)B_pT_A(\rb)\big[\N^p(x,\rr)+\N^p(x,\rr^{\prime})\big] \Big\},
  \label{amp_inc}
\end{eqnarray}
\end{widetext}
where $B_p$ is associated to the impact parameter profile function in the proton and $\N^p(x,\rr)$ is the dipole - proton scattering amplitude to be discussed in more detail in the next Section.

The functions 
$\Psi_{\gamma^*}(z,\,\rr)$ and $\Psi_{\gamma}(z,\,\rr)$ in the Eq. (\ref{totalcscoe1}) are the light-cone wave functions  of the virtual photon in the initial state and the real photon in the final state, respectively.  The 
variable $\rr$ defines the relative transverse
separation of the pair (dipole) and $z$ $(1-z)$ is the
longitudinal momentum fraction of the quark (antiquark). 
In the dipole formalism, the light-cone
 wave functions $\Psi(z,\,\rr)$ in the mixed
 representation $(r,z)$ are obtained through a two dimensional Fourier
 transform of the momentum space light-cone wave functions
 $\Psi(z,\,\rk)$.
The photon wave functions  are well known in literature \cite{kmw}. In the DVCS case, as one has a 
real photon at the final state, only the transversely polarized overlap function contributes 
to the cross section.  Summed over the quark helicities, for a given quark flavour $f$ it is 
given by \cite{kmw},
\begin{widetext}
\begin{eqnarray}
  (\Psi_{\gamma}^*\Psi)_{T}^f & = & \frac{N_c\,\alpha_{\mathrm{em}}
e_f^2}{2\pi^2}\left\{\left[z^2+\bar{z}^2\right]\varepsilon_1 K_1(\varepsilon_1 r) 
\varepsilon_2 K_1(\varepsilon_2 r) 
 +     m_f^2 K_0(\varepsilon_1 r) K_0(\varepsilon_2 r)\right\},
  \label{eq:overlap_dvcs}
\end{eqnarray}
\end{widetext}
where we have defined the quantities $\varepsilon_{1,2}^2 = z\bar{z}\,Q_{1,2}^2+m_f^2$ and 
$\bar{z}=(1-z)$. Accordingly, the photon virtualities are $Q_1^2=Q^2$ (incoming virtual 
photon) and $Q_2^2=0$ (outgoing real photon).

\section{QCD dynamics}
\label{qcd}

The   DVCS cross sections  in $eA$ collisions are expressed in terms of the  dipole-proton cross section $ \sigma_{dp}$, which in the eikonal approximation is given by:
\begin{equation} 
\sigma_{dp} (x, \rr) = 2 \int d^2 \rb \,  {\cal N}^p (x, \rr, \rb)\,\,,
\label{sdip}
\end{equation}
where  $\mathcal{N}^p(x,\rr,\rb)$ is  the imaginary part of the forward amplitude for the scattering between a small dipole
(a colorless quark-antiquark pair) and a dense hadron target, at a given
rapidity interval $Y=\ln(1/x)$. The dipole has transverse size given by the vector
$\rr=\rx - \ry$, where $\rx$ and $\ry$ are the transverse vectors for the quark
and antiquark, respectively, and impact parameter $\rb=(\rx+\ry)/2$.
At high energies the evolution with the rapidity $Y$ of
 $\mathcal{N}^p(x,\rr,\rb)$  is given in the Color Glass Condensate (CGC) formalism \cite{CGC} by the infinite hierarchy of equations, the so called
Balitsky-JIMWLK equations \cite{BAL,CGC}, which reduces in the mean field approximation to the Balitsky-Kovchegov (BK) equation \cite{BAL,kov} (For a detailed discussion about the subject see, e.g., Ref. \cite{alvioli}). 
In recent years,  the running coupling corrections to BK evolution kernel was explicitly calculated  \cite{kovwei1,balnlo},  including the  $\alpha_sN_f$ corrections to 
the kernel to all orders, and its solution studied in detail  \cite{javier_kov,javier_prl}. Basically, one has that the running of the coupling reduces the speed of the evolution to values compatible with experimental $ep$ HERA data \cite{bkrunning,weigert,aamqs}.  The numerical solutions of the running coupling BK equation presented in Refs. \cite{bkrunning,weigert,aamqs} assumed the translational invariance approximation, which implies  $\mathcal{N}^p(x,\rr,\rb) = \mathcal{N}^p(x,\rr) S(\rb)$ and $\sigma_{dp} (x, \rr) = \sigma_0 \cdot  {\cal N}^p (x, \rr)$, with the normalization of the dipole cross section ($\sigma_0$) being fitted to data.  It is important to emphasize that the normalization $\sigma_0$  defines $B_p$ in Eq. (\ref{amp_inc}), since they are related: $B_p = \sigma_0 / 4 \pi$. Moreover, ${\cal N}^p (x, \rr)$ is the main input for the calculation of the incoherent DVCS cross section [See Eq. (\ref{amp_inc})].    
In what follows we will consider the solutions for the scattering amplitude obtained in Refs. \cite{bkrunning,aamqs}, denoted respectively by rcBK and AAMQS, which differ in the set of experimental data used to constrain the initial conditions and in the treatment of the heavy quark contributions. While in the rcBK model \cite{bkrunning} the experimental data for the inclusive and longitudinal proton structure functions measured in $ep$ collisions at HERA were considered in the fit,  the more precise HERA data for the reduced cross sections were used 
in Ref.  \cite{aamqs}. Moreover, in the rcBK model the contribution of the heavy quarks was disregarded. In contrast, in Ref. \cite{aamqs} the authors have performed a non-linear QCD analysis of new HERA data at small-$x$ including heavy quarks. We denote the corresponding solution of the running coupling BK equation by AAMQS(H). For comparison, we  also will consider the solution derived in Ref. \cite{aamqs}, denoted  AAMQS(L) hereafter,  which was obtained taking into account only light quarks (See Ref. \cite{aamqs} for more details). 

A shortcoming of the solutions presented in Refs.  \cite{bkrunning,weigert,aamqs} is that they disregard the impact parameter dependence. Unfortunately, impact-parameter dependent numerical solutions to the BK equation are very difficult to obtain \cite{stasto}. Moreover, the choice of the impact-parameter profile of the dipole amplitude entails intrinsically nonperturbative physics, which is beyond the QCD weak coupling approach of the BK equation. In fact, the BK equation  generates a power law Coulomb-like tail, which is not confining at large distances and therefore can violate the Froissart bound \cite{kovner}.  As demonstrated in Refs. \cite{stasto,ikeda}, confinement can be modelled by the modification of the dipole evolution kernel of the BK equation  by including an effective gluon mass, which implies that the Froissart bound limit is satisfied. An alternative is to construct phenomenological models which incorporate the expected $b$-dependence of the scattering amplitude. Although a complete analytical solution of the BK equation is still lacking, its main properties at fixed $\rb$ are known: (a) for the interaction of a small dipole ($r = |\rr| \ll
1/Q_s$), $\mathcal{N}^p(x,\rr,\rb) \approx \rr^2$, implying  that
this system is weakly interacting; (b) for a large dipole ($r \gg
1/Q_s$), the system is strongly absorbed and therefore
$\mathcal{N}^p(x,\rr,\rb) \approx 1$.  This property is associated  to the
large density of saturated gluons in the hadron wave function. In the last years, several groups have constructed phenomenological models which satisfy the asymptotic behaviours of the BK equation in order to fit the HERA and RHIC data (See e.g. Refs. \cite{GBW,iim,kkt,dhj,Goncalves:2006yt,buw,kmw}). 
In particular, in Ref. \cite{kmw} the authors have proposed  a generalization of Iancu - Itakura - Munier model \cite{iim}, which is based on the CGC formalism, with  the inclusion of   the impact parameter dependence in the dipole - proton scattering amplitude (denoted bCGC  hereafter).  In the bCGC model the dipole - proton scattering amplitude is given by \cite{kmw} 
\begin{widetext}
\begin{eqnarray}
\mathcal{N}^p(x,\rr,\rb) =   
\left\{ \begin{array}{ll} 
{\mathcal N}_0\, \left(\frac{ r \, Q_s(b)}{2}\right)^{2\left(\gamma_s + 
\frac{\ln (2/r Q_s(b))}{\kappa \,\lambda \,Y}\right)}  & \mbox{$r Q_s(b) \le 2$} \\
 1 - \exp^{-A\,\ln^2\,(B \, r  Q_s(b))}   & \mbox{$r Q_s(b)  > 2$} 
\end{array} \right.
\label{eq:bcgc}
\end{eqnarray}
\end{widetext}
with  $\kappa = \chi''(\gamma_s)/\chi'(\gamma_s)$, where $\chi$ is the 
LO BFKL characteristic function.  The coefficients $A$ and $B$  
are determined uniquely from the condition that $\mathcal{N}^p(x,\rr,\rb)$, and its derivative 
with respect to $r\,Q_s(b)$, are continuous at $r\,Q_s(b)=2$. 
In this model, the proton saturation scale $Q_s(b)$ depends on the impact parameter:
\begin{equation} 
  Q_s(b)\equiv Q_s(x,b)=\left(\frac{x_0}{x}\right)^{\frac{\lambda}{2}}\;
\left[\exp\left(-\frac{{b}^2}{2B_{\rm CGC}}\right)\right]^{\frac{1}{2\gamma_s}}.
\label{newqs}
\end{equation}
The parameter $B_{\rm CGC}$  was  adjusted to give a good 
description of the $t$-dependence of exclusive $J/\psi$ photoproduction.  
Moreover, the factors $\mathcal{N}_0$ and  $\gamma_s$  were  taken  to be free. In this way a very good description of  $F_2$ data was obtained. 
Recently, this model has been improved by the fitting of its free parameters considering the high precision combined HERA data \cite{amir}. The set of parameters  which will be used here are the following: $\gamma_s = 0.6599$, $\kappa = 9.9$, $B_{CGC} = 5.5$ GeV$^{-2}$, $\mathcal{N}_0 = 0.3358$, $x_0 = 0.00105$ and $\lambda = 0.2063$.  
In order to estimate the model dependence of our predictions, in what follows we also will present the predictions obtained considering  the phenomenological models discussed in Refs. \cite{GBW,iims},  which  assume the factorization of the impact parameter dependence and propose different forms for  $\mathcal{N}^p(x,\rr)$. In particular, the GBW model \cite{GBW} assume that 
 $\mathcal{N}^p(x,\rr)=1-e^{-\rr^2Q_{s}^2(Y)/4}$ and  the  IIMS model \cite{iim,iims} can be obtained from Eq. (\ref{eq:bcgc}) disregarding the impact parameter dependence of the saturation scale. For the GBW model we use the following set of parameters \cite{GBW}: $x_0=3 \times 10^{-4}$, $\lambda=0.288$ and $\sigma_0=23.03$ mb.
 For the IIMS model we assume  that \cite{iims} $\N_0 =0.7$, $\gamma_s = 0.7376$, $\kappa =9.9$, $\lambda = 0.2197$, $x_0=1.632 \times 10^{-5}$ and $\sigma_0=27.28$ mb.

\begin{figure}[t]
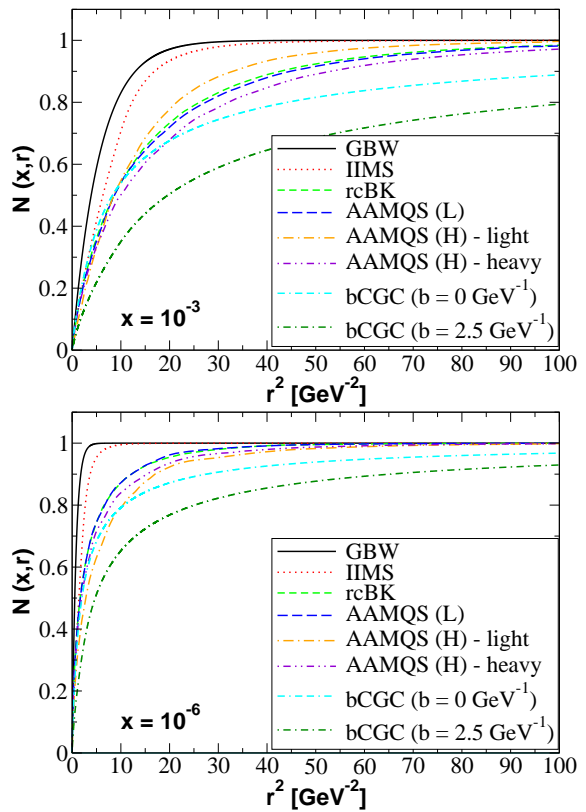

\begin{center}
\scalebox{0.3}{\includegraphics{N_X3.eps}}
\scalebox{0.3}{\includegraphics{N_X6.eps}}
\caption{(Color online) Dependence of the  dipole - proton scattering amplitude  in the squared pair separation $\rr^2$ at 
different values of $x$.}
\label{Dipole}
\end{center}
\end{figure}

Before presenting our results for the nuclear DVCS cross sections in the next Section, we compare the dipole - proton scattering amplitudes predicted by the solutions of the running coupling BK equation (rcBK and AAMQS)  with those from phenomenological models (IIMS and GBW). In Fig. \ref{Dipole} we analyse  the pair separation dependence of 
the dipole scattering amplitude $\mathcal{N}^p$ for two different values of $x$. For comparison we also present the predictions of the bCGC for two distinct values of the impact parameter. It is important to emphasize that in Ref. \cite{aamqs}, when the heavy quark contribution is included in the analysis, the authors have considered different initial values of the parameters in the initial condition for light and heavy quarks. It implies two different parametrizations for the AAMQS(H) solution, one associated to light  and other to heavy quarks. It explains the presence of two AAMQS(H) curves in Fig. \ref{Dipole}. 
We observe that while the GBW and IIM  
parametrizations present a similar behaviour for small $\rr^2$, the rcBK and AAMQS one predict 
a smoother dependence. These differences are associated to the distinct behaviours of the linear regime assumed by the models. In the saturation regime (large $\rr^2$), the  GBW and IIM   parametrizations saturates 
for large pair separations,  while the rcBK and AAMQS  one still present a residual dependence, 
demonstrating that the asymptotic regime is only reached for very large pair separations.  The delayed saturation of the  scattering amplitude at small values of $x$ is still more pronounced in the bCGC predictions.
The characteristic feature which is evident in the GBW and  IIMS models is that the dipole scattering amplitude saturates for smaller dipoles when $x$ assumes  smaller values, which is directly associated to the energy dependence of the proton saturation scale, which defines the onset of the saturation regime. Finally, the different models presented in Fig. \ref{Dipole} predict very distinct behaviours for the transition region between small and large pair separations. As we will demonstrate in the next Section, these differences has strong influence on the energy behaviour of the DVCS cross section.

\begin{figure}[t]
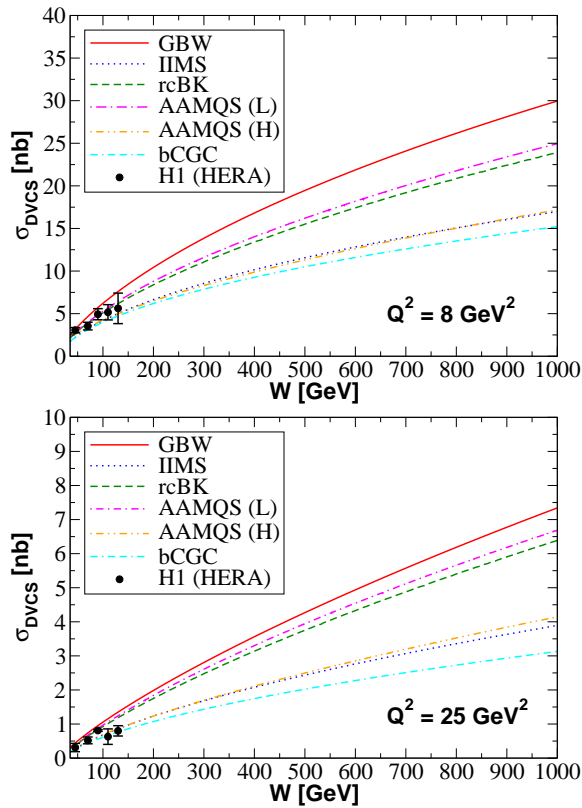

\begin{center}
\scalebox{0.3}{\includegraphics{Q8.eps}}
\scalebox{0.3}{\includegraphics{Q25.eps}}
\caption{(Color online) Energy dependence of the DVCS cross section in $ep$ collisions for two values of the photon virtuality $Q^2$. Data from H1 Collaboration \cite{h1}.}
\label{ep}
\end{center}
\end{figure}

\begin{figure}[t]
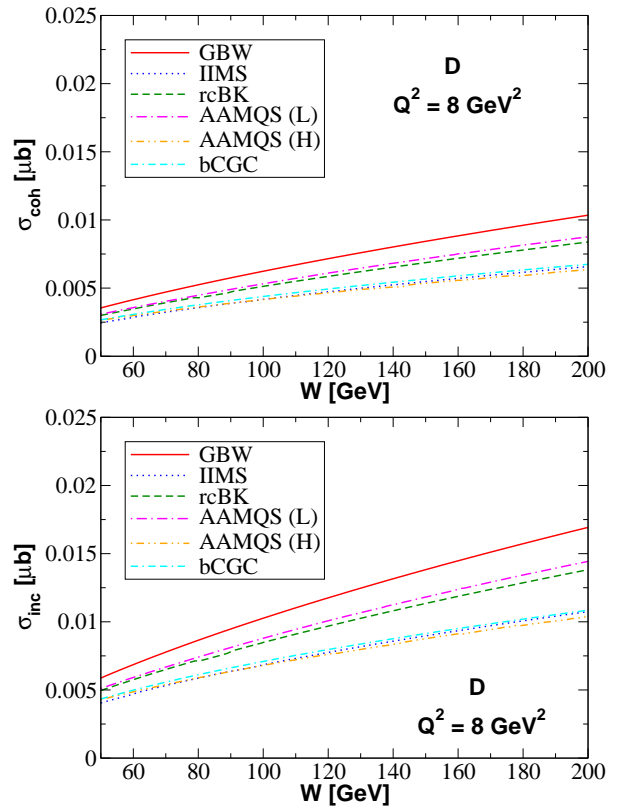

\begin{center}
\scalebox{0.3}{\includegraphics{coherent_q8_d.eps}}
\scalebox{0.3}{\includegraphics{incoherent_q8_d.eps}}
\caption{(Color online) Energy dependence of the coherent (upper panel) and incoherent (lower panel) DVCS cross sections for $A = 2$ (D) and $Q^2 = 8$ GeV$^2$.}
\label{deuterio}
\end{center}
\end{figure}

\begin{figure}[t]
\begin{center}
\scalebox{0.3}{\includegraphics{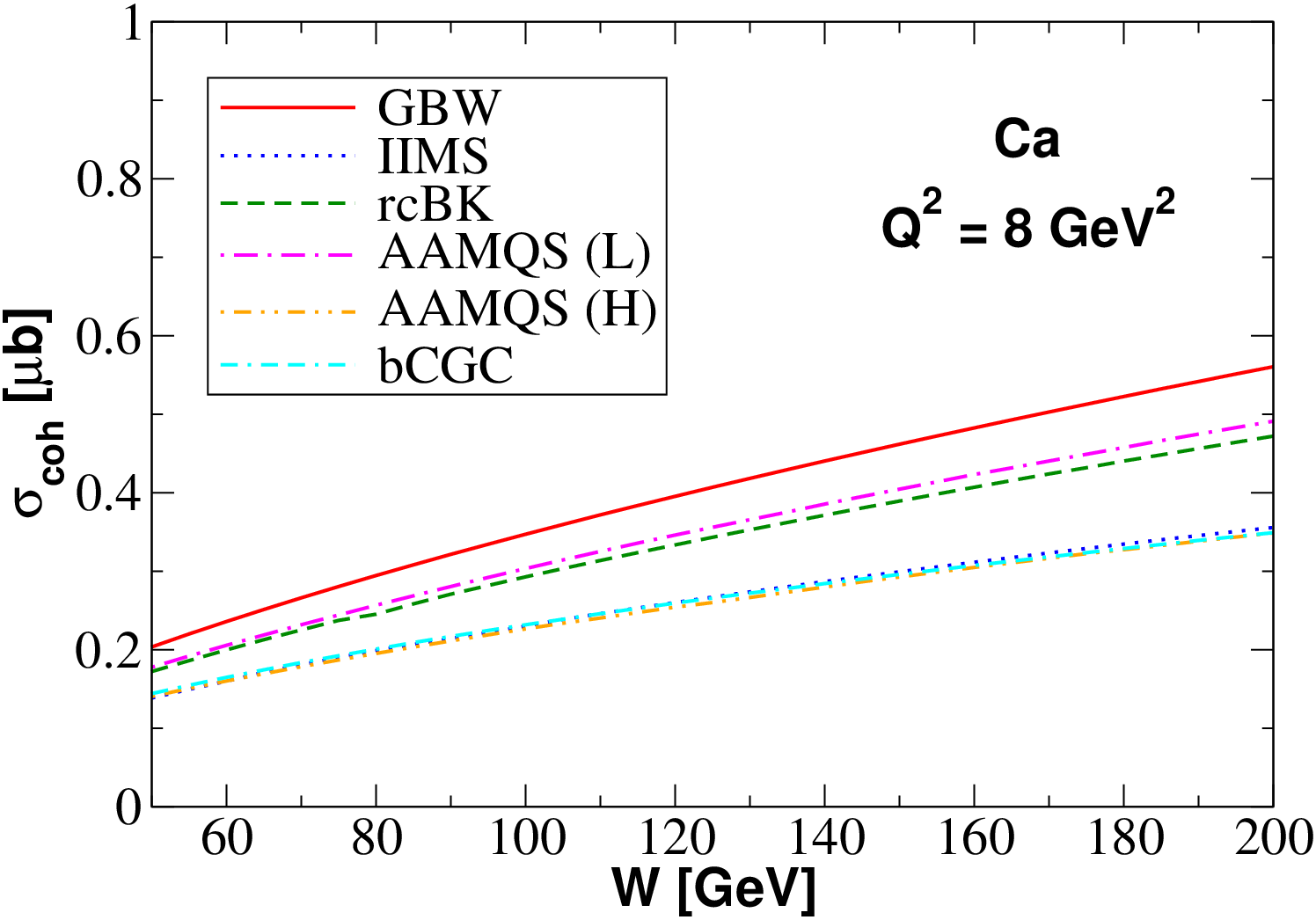}}
\scalebox{0.3}{\includegraphics{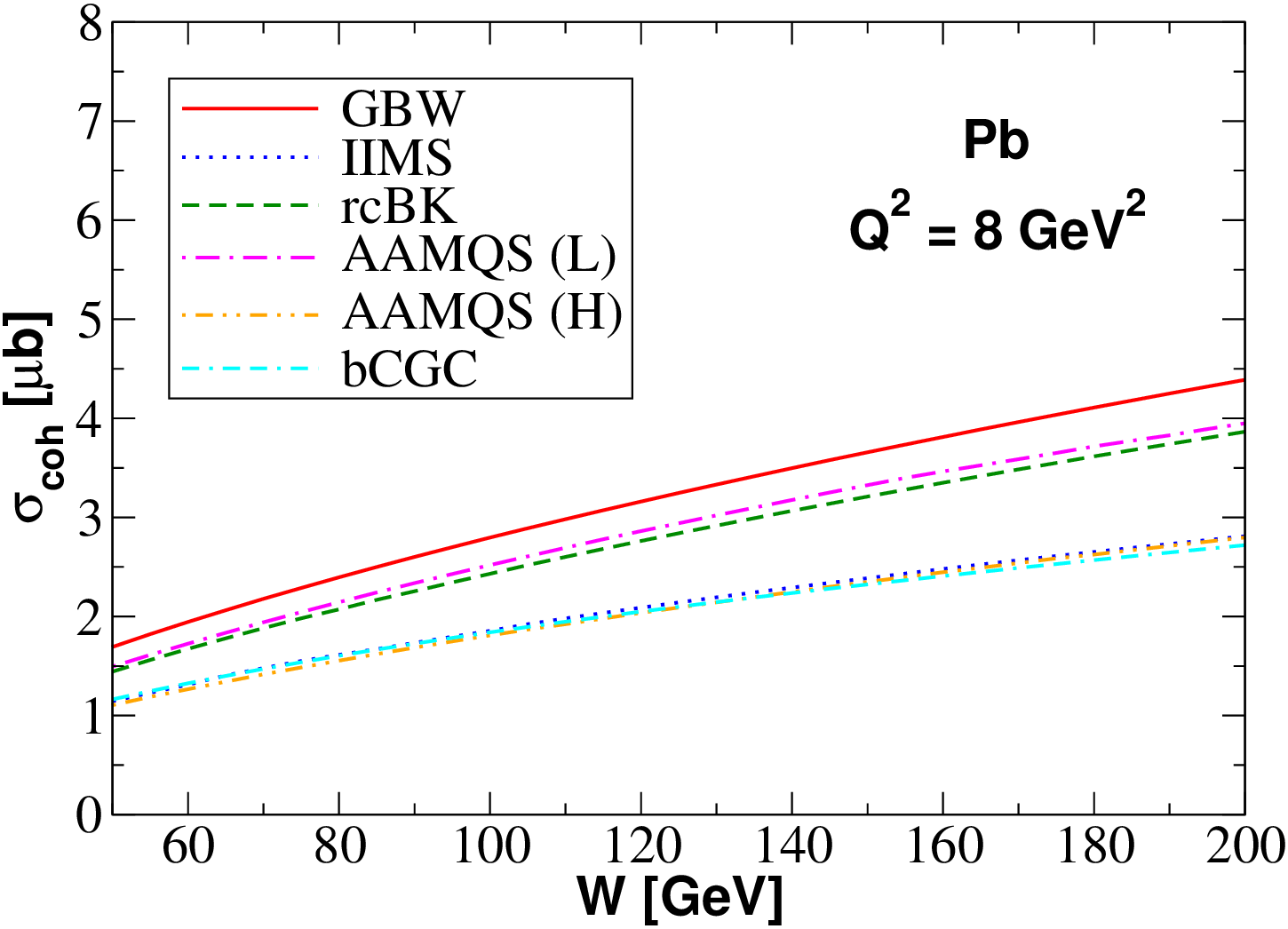}}
\caption{(Color online) Energy dependence of the coherent DVCS cross section for two different nuclei and $Q^2 = 8$ GeV$^2$.}
\label{coh_energy}
\end{center}
\end{figure}

\begin{figure}[t]
\begin{center}
\scalebox{0.3}{\includegraphics{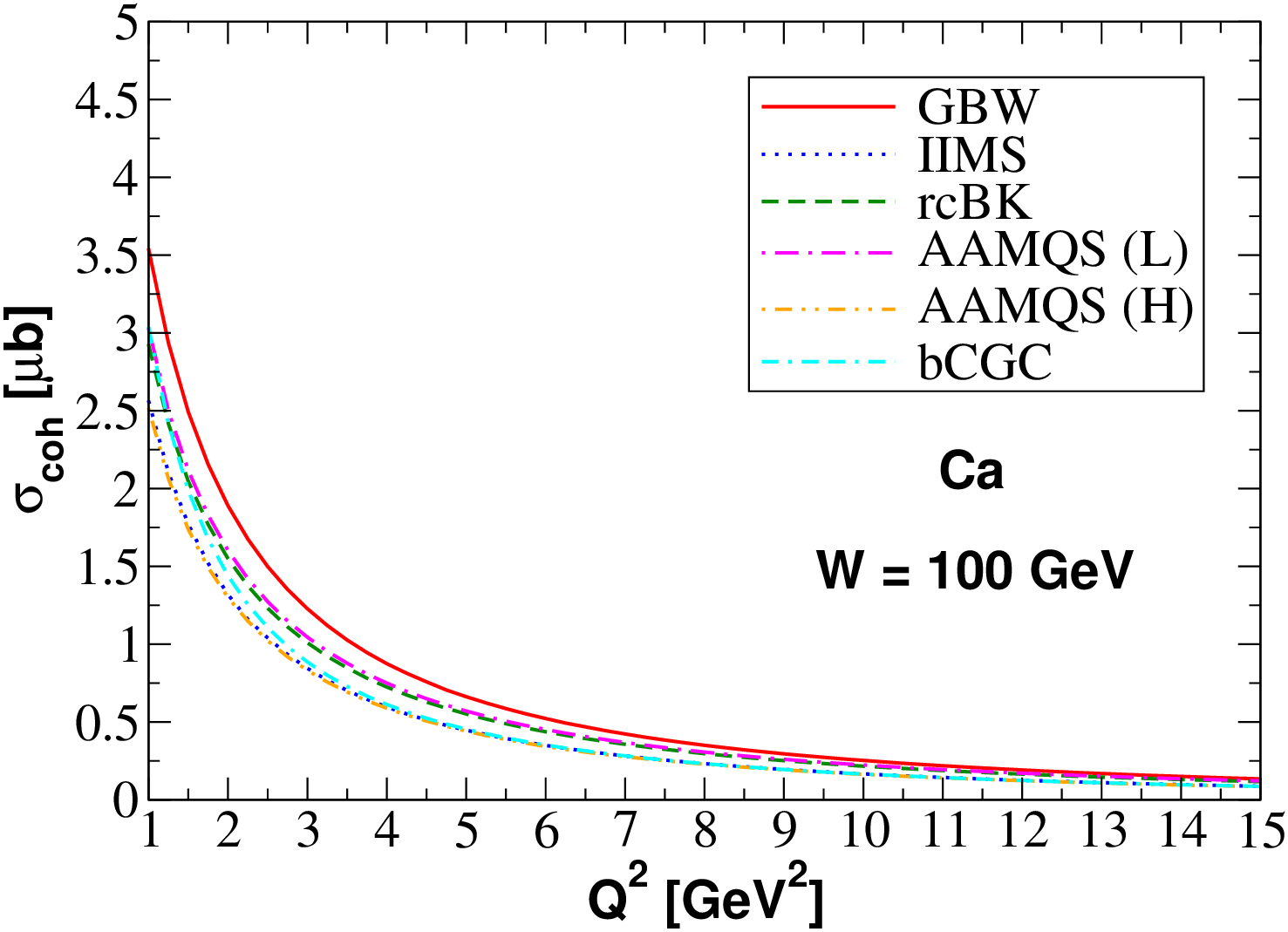}}
\scalebox{0.3}{\includegraphics{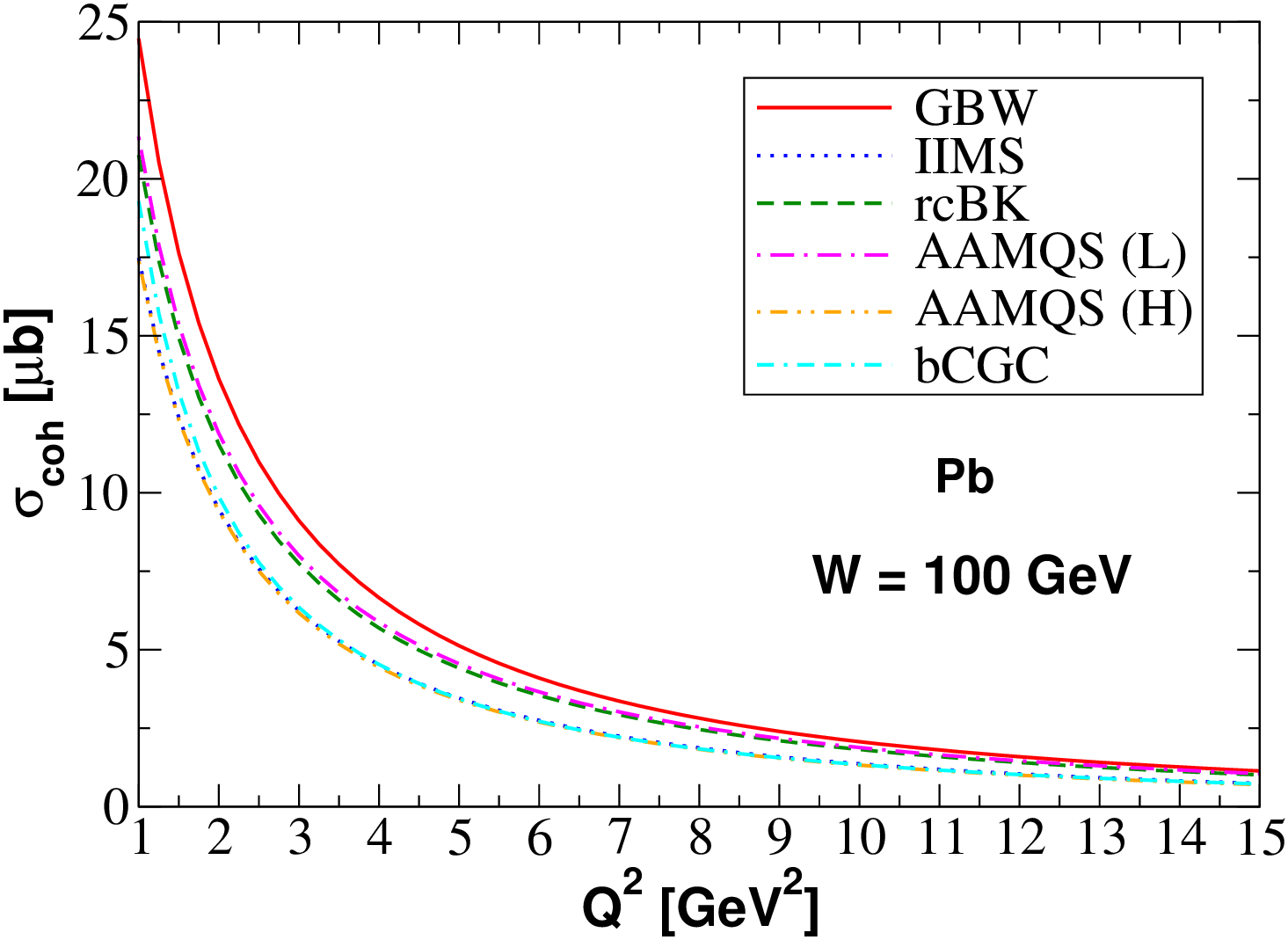}}
\caption{(Color online) Dependence on the photon virtuality $Q^2$ of the coherent DVCS cross section for two different nuclei and $W = 100$ GeV.}
\label{coh_q2}
\end{center}
\end{figure}

\begin{figure}[t]
\begin{center}
\scalebox{0.3}{\includegraphics{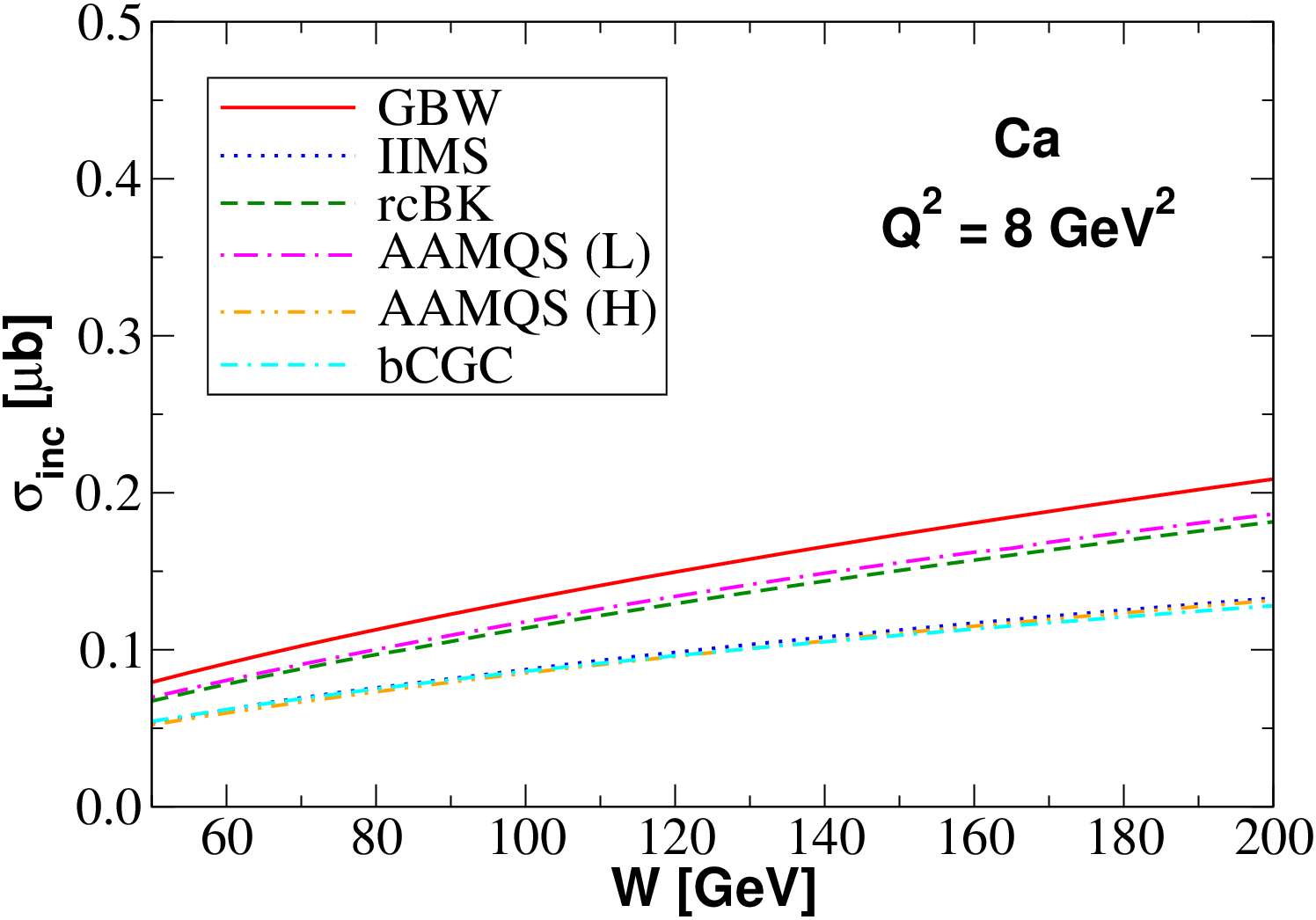}}
\scalebox{0.3}{\includegraphics{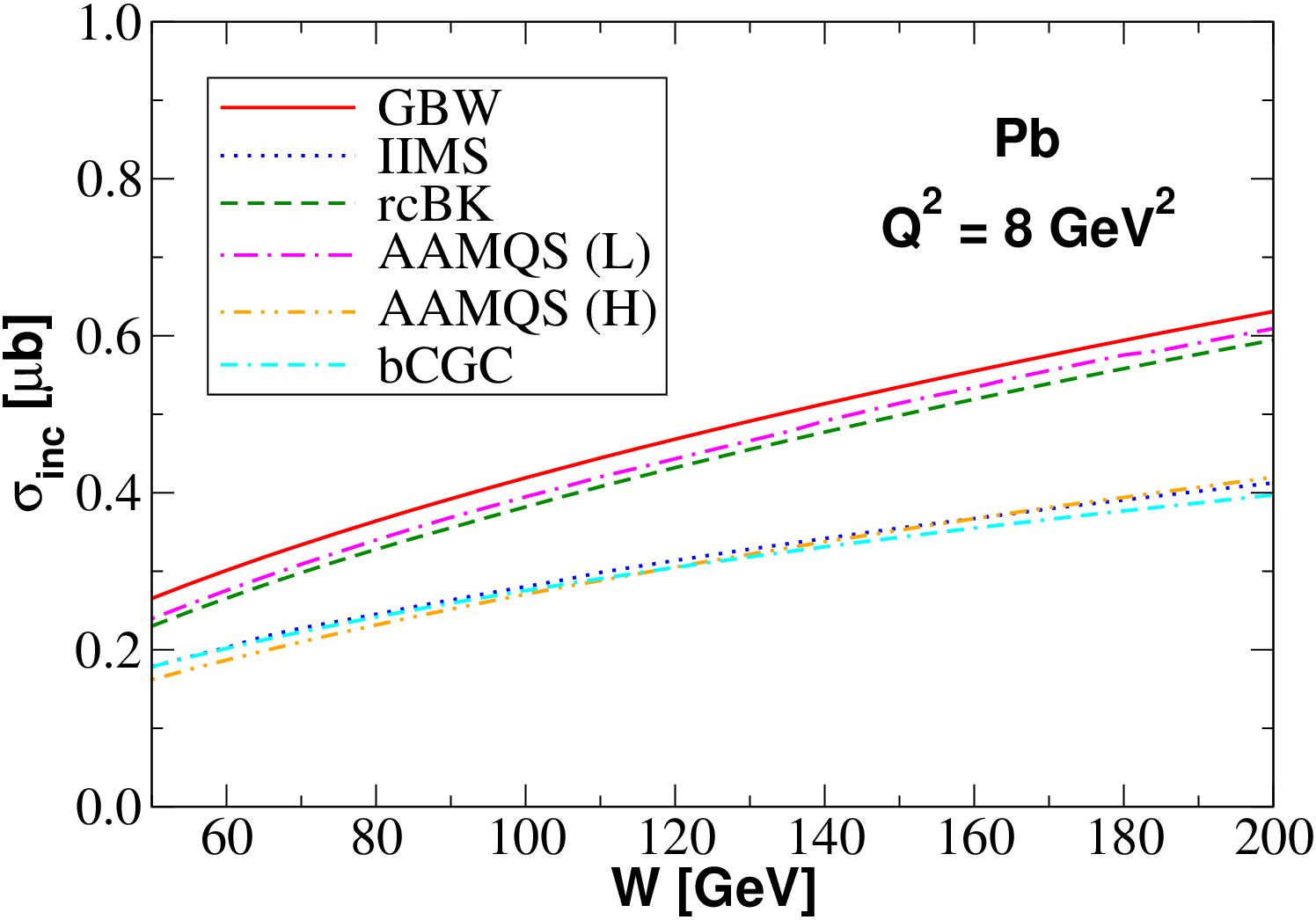}}
\caption{(Color online) Energy dependence of the incoherent DVCS cross section for two different nuclei and $Q^2 = 8$ GeV$^2$.}
\label{inc_energy}
\end{center}
\end{figure}

\begin{figure}[t]
\begin{center}
\scalebox{0.3}{\includegraphics{incoherent_w100_ca.eps}}
\scalebox{0.3}{\includegraphics{incoherent_w100_pb.eps}}
\caption{(Color online) Dependence on the photon virtuality $Q^2$ of the coherent DVCS cross section for two different nuclei and $W = 100$ GeV.}
\label{inc_q2}
\end{center}
\end{figure}


\begin{figure}[t]
\begin{center}
\scalebox{0.3}{\includegraphics{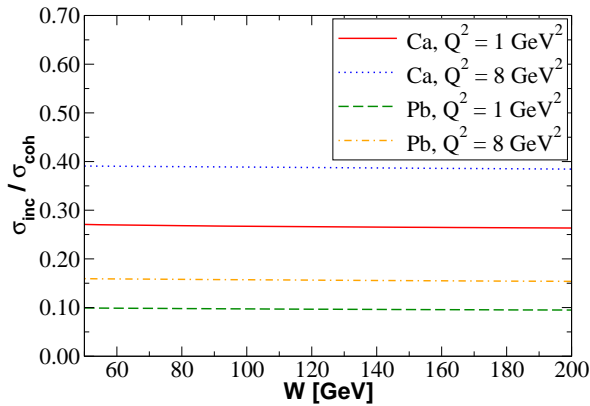}}
\caption{(Color online) Energy dependence of the ratio between the incoherent and coherent cross sections obtained using the rcBK model for the dipole - proton scattering amplitude for different nuclei and values of $Q^2$.}
\label{ratio2}
\end{center}
\end{figure}

\begin{figure}[t]
\begin{center}
\scalebox{0.3}{\includegraphics{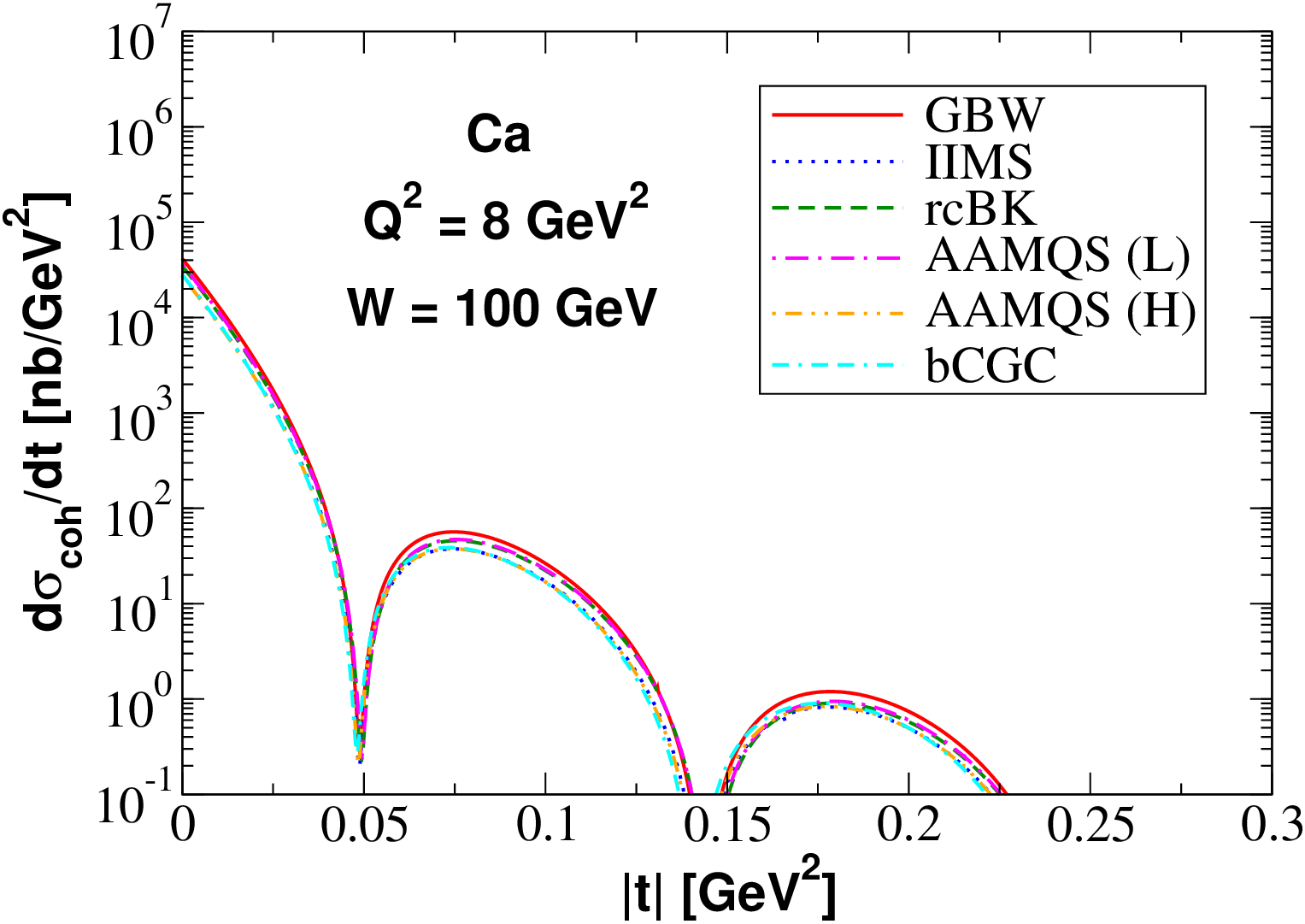}}
\scalebox{0.3}{\includegraphics{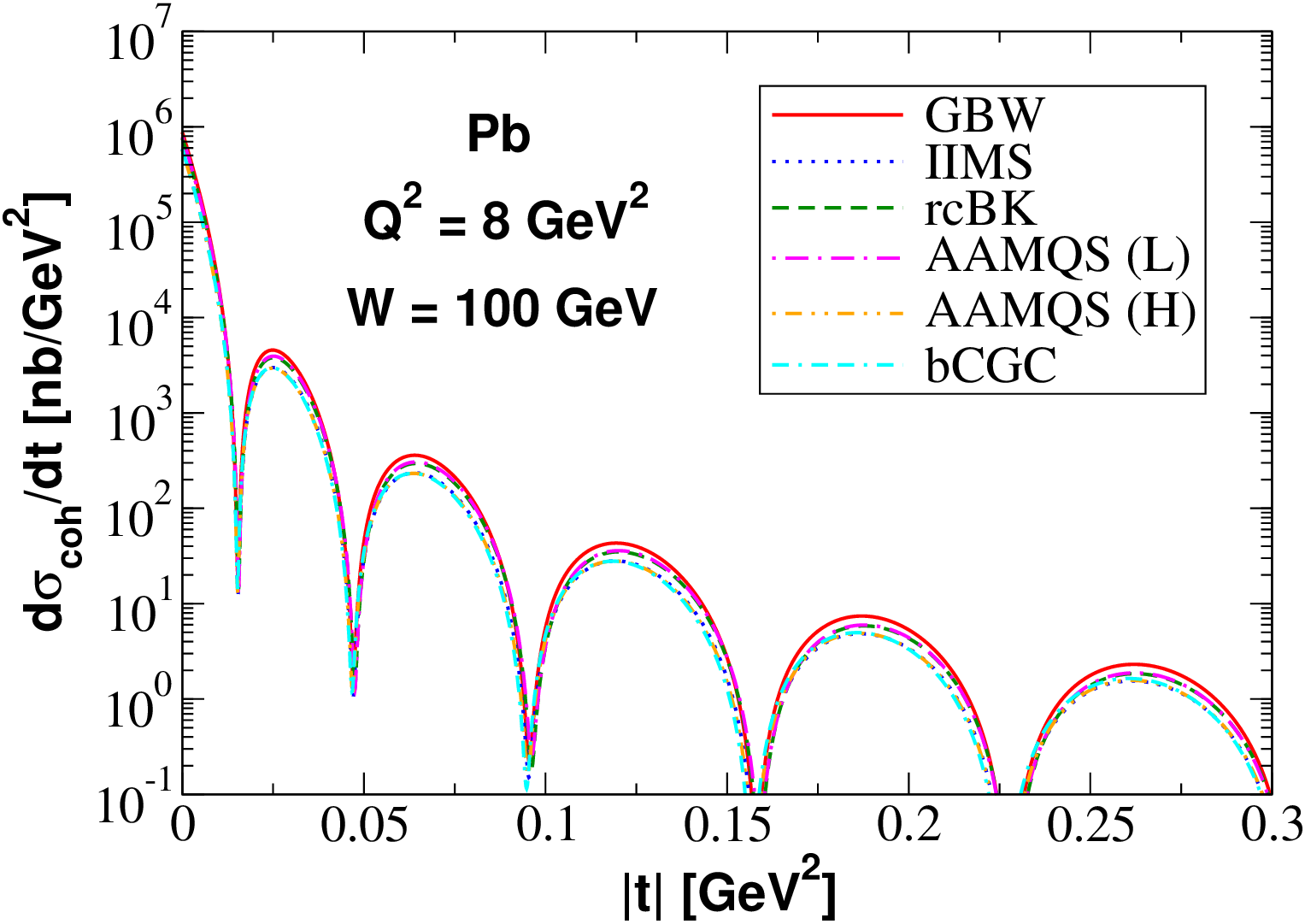}}
\caption{(Color online) Differential cross section for coherent interactions.}
\label{coh_t}
\end{center}
\end{figure}

\begin{figure}[t]
\begin{center}
\scalebox{0.3}{\includegraphics{t_inc_q8_w100_ca.eps}}
\scalebox{0.3}{\includegraphics{t_inc_q8_w100_pb.eps}}
\caption{ (Color online) Differential cross section for incoherent interactions. }
\label{inc_t}
\end{center}
\end{figure}

\begin{figure}[t]
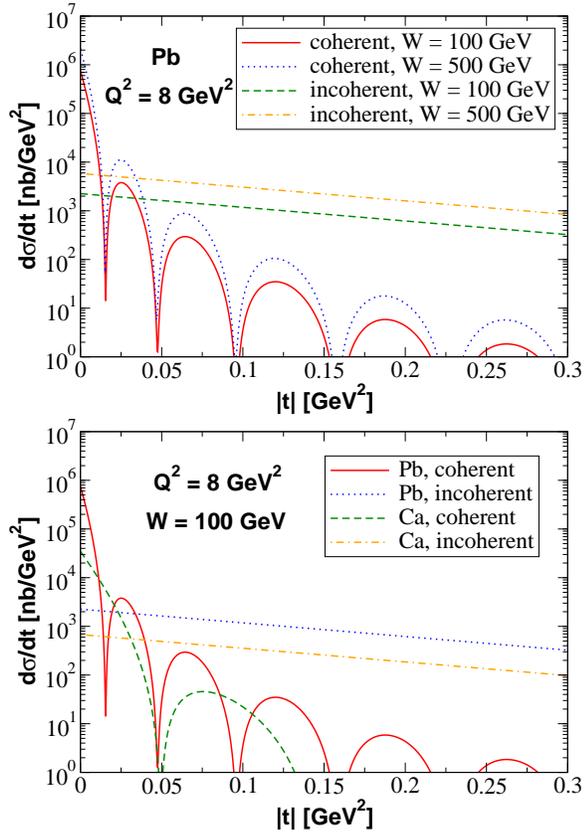

\begin{center}
\scalebox{0.3}{\includegraphics{t_coh_inc_q8_pb.eps}}
\scalebox{0.3}{\includegraphics{t_coh_inc_q8_w100_pb_ca.eps}}
\caption{(Color online) Comparison between the coherent and incoherent cross sections for different energies (upper panel) and distinct nuclei (lower panel).}
\label{t_coh_inc}
\end{center}
\end{figure}

\section{Results}
\label{res}

We initially  estimate the energy dependence of the  DVCS cross section in $ep$ collisions. As in Ref. \cite{vic_anelise} we take into account the corrections associated to the  skewness factor $R_g$ and to the real part of the scattering amplitude. Basically, the expressions for the DVCS cross sections presented in Section \ref{nucDVCS} have been multiplied by the factor     $R_g^2(1 + \beta^2)$, with the factor $R_g$ taking into account that the gluons emitted from the quark and antiquark into the dipole carry different momentum fractions, and the factor $\beta$ being the ratio of real to  imaginary parts of the scattering amplitude (For details see, e.g. Ref. \cite{vic_anelise}). Moreover, when using as input in our $ep$ calculations the dipole models which disregard the impact parameter dependence (rcBK, AAMQS, GBW and IIMS), we will assume that 
$\sigma = \frac{1}{B} \cdot \frac{d\sigma}{dt}|_{t = 0}$, with the  
slope parameter $B$ being given by the experimental parametrization proposed in Ref. \cite{h1}: 
$B\,(Q^2)=a[1-b\log(Q^2/Q_0^2)]$, with $a=6.98 \pm 0.54 $ GeV$^2$, $b=0.12 \pm 0.03$ and 
$Q_0^2 = 2$ GeV$^2$. 
In contrast, in the case of the bCGC model, the cross section will be calculated by the integration of the differential cross section $\frac{d\sigma}{dt}$ as given in Eq. (4) of the Ref. \cite{vic_anelise}. In Fig. \ref{ep} we present our predictions for the DVCS cross section in $ep$ collisions
as a function of the photon-target c.m.s energy, $W$, for two values of the photon virtuality: $Q^2=8$ and 25 GeV$^2$. For comparison the HERA data from the H1 Collaboration \cite{h1} are also presented. 
We have that the GBW predictions are an upper bound for the cross section, while the bCGC predictions can be considered a lower bound. The rcBK and AAMQS(L) predictions are similar, which is expected, since they are based on similar assumptions. In contrast, the AAMQS(H) prediction is similar to the IIMS and bCGC predictions.
 In the HERA kinematical regime, the difference between the predictions  is not large, but becomes a factor 2 in the range which will be probed by the future $ep$ colliders \cite{LHeC}. This huge difference is directly associated to the distinct behaviours for the dipole scattering amplitude in the transition region between small and large dipoles observed in Fig. \ref{Dipole}. We have that the description of the experimental data by the different models is satisfactory, in particular at $Q^2 = 8$ GeV$^2$, where we expect a larger contribution of the saturation effects. In contrast, for large values of $Q^2$, the main contribution comes from smaller dipoles, which implies that we are probing the linear regime. In this regime, effects associated to the evolution in $Q^2$, which are not included in the dipole models considered in this paper, can contribute. Consequently, in what follows we will restrict our analysis of DVCS cross sections to values of   $Q^2$ smaller than 15 GeV$^2$.

Lets now estimate the coherent and incoherent DVCS cross sections for  different values of the nuclear mass number: $A = 2$ (D),  $A = 40$ (Ca) and 208 (Pb).  These results are an update of  those presented in Ref. \cite{vic_erike} corresponding  to  AAMQS predictions and light nuclei. For $A = 2$, the {\it oomph} factor $A^{\frac{1}{3}}$ is close to one, implying a small enhancement of non-linear effects in comparison to $ep$ collisions. However, the study of coherent and incoherent interactions in $eD$ collisions can be useful to test the color dipole formalism, as well as to compare its predictions with those obtained using very distinct frameworks (See, e.g. Refs. \cite{scopetta,siddikov}). For completeness, in Fig.  \ref{deuterio} we present the predictions   for the energy dependence of the coherent and incoherent $eD$ cross sections considering as input the different models for the dipole - proton scattering amplitude discussed before. On the other hand, for $A = 40$ (Ca) and 208 (Pb) we expect the enhancement of the non-linear effects, which makes the study of the DVCS process useful to constrain the magnitude of these effects. Consequently, considering the main goal of this paper,  in what follows we will restrict our analysis to these two nuclei.  In Fig. \ref{coh_energy} we present our predictions for the energy dependence of the coherent cross sections for a fixed virtuality and in Fig. \ref{coh_q2} the $Q^2$ dependence  for a fixed energy, $W=100$ GeV. Our results for $A = 40$ and 208  agree with the bCGC and rcBK predictions presented in Ref. \cite{vic_erike}. We obtain that the cross sections increase with the energy and nuclear mass number and decrease with the photon virtuality. As expected from Fig. \ref{Dipole}, the rcBK and AAMQS(L) predictions are very similar in the energy range analysed ($W \le 200$ GeV). At $Q^2 = 8$ GeV the bCGC, IIMS and AAMQS(H) are almost identical. However, as can be verified in the Fig. \ref{coh_q2}, its predictions differ at smaller values of $Q^2$, due to the different behaviour predict by these models in the transition region between small and large dipoles. In the energy range which will be probed in future $eA$ colliders ($W \le 140$ GeV), the two AAMQS predictions differ by $\approx 50 \%$. These conclusions are also valid for the  incoherent cross sections presented in  Figs. \ref{inc_energy} and \ref{inc_q2}. In Fig. \ref{ratio2} we present the energy dependence of the ratio between the incoherent and coherent cross sections for different nuclei and $Q^2$ using the  rcBK model. We obtain that the ratio decreases at smaller values of  $Q^2$ and increases at smaller values of $A$. It is important to emphasize that  we have verified that our predictions are  almost independent of the model used for the dipole - proton scattering amplitude. Finally, we have obtained that the $A$ dependence of our predictions can be approximated by $\sigma_{coh} \propto A^{1.3}$ and $\sigma_{inc} \propto A^{0.75}$ for $Q^2 = 8$ GeV$^2$ and $W = 120$ GeV.

Lets now extend our analysis for the differential distributions $d\sigma/dt$ for coherent and incoherent interactions. In Fig. \ref{coh_t} we present our predictions for the coherent DVCS cross section for different nuclei and $Q^2 = 8$ GeV$^2$.
We obtain that the coherent cross section clearly exhibits the typical diffractive pattern, with  the dips in the range $|t| \le 0.3$ GeV$^2$  increasing with the mass atomic number. Moreover, the positions of the dips are almost independent of the dipole - proton model used as input in the calculations. In Fig. \ref{inc_t} we present our predictions for the incoherent cross section. In this case no dip is present and the predictions of the different models being similar at large $t$. A more detailed comparison between the coherent and incoherent cross sections is presented in Fig. \ref{t_coh_inc}, where we consider distinct values of the energy (upper panel) and  different nuclei (lower panel).  As expected from the studies of the exclusive vector meson production \cite{Caldwell,Lappi_inc,Toll}, we obtain that coherent DVCS cross section dominates at small $t$ ($|t|\cdot R_A^2/3 \ll 1$), the signature being a sharp 
forward diffraction peak. On the other hand, incoherent DVCS production  dominates  
at large $t$ ($|t|\cdot R_A^2/3 \gg 1$), the $t$-dependence being to a good 
accuracy the same as in the production off  free nucleons. We obtain that the magnitude of the differential cross sections at $|t| = 0$ increases with the energy and the nuclear mass number, as expected from the analysis of the total cross sections, and that the positions of the dips are almost energy independent.

\section{Conclusions}
\label{conc}

The study of the hadronic structure using hard exclusive reactions in deep inelastic scattering (DIS) processes, such as deeply virtual Compton scattering (DVCS) and vector meson production, have been one of the main focuses of hadronic physics in the last years. These processes have been the subject of intensive theoretical and experimental investigations. One of the main motivations for these studies is the possibility to probe  the QCD dynamics at high energies,  driven by the gluon content of the target (proton or nucleus) which is strongly subject to non-linear effects (parton saturation) effects. As in $eA$ collisions the coherent contributions from many nucleons effectively amplify the gluon density probed, the study of exclusive observables in the future electron - ion colliders is the ideal scenario to constrain the QCD dynamics at high energies. In this paper we have presented a detailed study of the nuclear DVCS process, which complement the existing analysis about vector meson production in $eA$ collisions. Although the DVCS cross section is smaller than the vector meson one, it is not affected by the theoretical uncertainties associated to the scarce knowledge of the vector meson wave functions. Consequently, the DVCS process can be considered a direct probe of the QCD dynamics which describes the dipole - target interaction. We have assumed in our studies that the dipole - nucleus scattering amplitude can be described in terms of   the dipole - proton one and considered different models for $\mathcal{N}^p(x,\rr)$. In particular, we have extended the results obtained in Ref. \cite{vic_erike}, presenting  the first time the predictions of the AAMQS model for the DVCS process. The coherent and incoherent cross sections were estimated for different nuclei, energies and photon virtualities. We have verified that the incoherent contribution increases with the photon virtuality and decreases with the nuclear mass number. Moreover, we estimated the differential cross sections for the nuclear DVCS and have analysed in detail the dependence on the squared momentum transfer. As for vector meson production, we have obtained that the coherent cross section dominates at small $t$ and the incoherent one at large $t$. We have demonstrated that the number of dips at small $t$ increases with the atomic number, with the position of the dips being almost   independent of the model used to treat the dipole - proton interaction and the center-of-mass energy. Both results are robust predictions from the saturation physics, which can be used to challenge the treatment of the non-linear QCD dynamics in the kinematical range of future electron - ion experiments.


\begin{acknowledgements}
This work was  partially financed by the Brazilian funding
agencies CNPq, CAPES and FAPERGS.
\end{acknowledgements}

\end{document}